\documentclass[twocolumn,floats,showpacs,superscriptaddress,pre]{revtex4}
\usepackage{amsmath,amssymb,bm}
\usepackage{graphics}
\usepackage{epsfig}
\usepackage{graphicx}

\begin{document}

\title{Quantum Action Principle in Relativistic Mechanics}
\author{Natalia Gorobey and Alexander Lukyanenko}
\email{alex.lukyan@rambler.ru}
\affiliation{Department of Experimental Physics, St. Petersburg State Polytechnical
University, Polytekhnicheskaya 29, 195251, St. Petersburg, Russia}

\begin{abstract}
A quantum version of the action principle is considered in the case of a
free relativistic particle. The classical limit of the quantum action is
obtained.
\end{abstract}

\maketitle
\date{\today }





\section{\textbf{INTRODUCTION}}

In the works \cite{GL,GL1} a new form of non-relativistic quantum
mechanics in terms of a quantum action principle was proposed. The quantum
action principle was formulated for a new object - a wave functional $\Psi %
\left[ x\left( t\right) \right] $ which, unlike a wave function $\psi
\left( x,t\right) $, describes dynamics of a particle as a movement
along a trajectory $x\left( t\right) $. The wave functional has the meaning
of a probability density in the space of trajectories. It is this
description of dynamics that is most appropriate for relativistic quantum
mechanics. In the relativistic mechanics a trajectory of a particle must
be replaced by an invariant geometrical object - a world line, $x^{\mu
}\left( \tau \right)$, where $\mu =0,1,2,3$ in the Minkowsky space. Here $\tau $ is
an arbitrary parameter along the world line. As a result, the wave
functional $\Psi \left[ x^{\mu }\left( \tau \right) \right] $ becomes
relativistic invariant preserving its probabilistic interpretation.

A special feature of relativistic mechanics is the presence of an additional
symmetry - an independence of the action on the parametrization of a world
line of a particle (see, for example, \cite{W}). In ordinary quantum
mechanics based on a wave function,
it is necessary for its probabilistic interpretation to fix a time parameter by use of an additional gauge condition.
In our approach the re-parametrization
invariance must be unbroken. Gauge parameters which ensure invariance of
the action must be added to a set of variational parameters of the
quantum action principle. Therefore, the advantage of the new approach is
the possibility of probabilistic interpretation of the quantum theory without
a loss of its covariance. In the present paper the quantum action principle is
considered in the case of a free relativistic particle. The classical limit
of the quantum action is obtained.

\section{QUANTUM ACTION PRINCIPLE}

We begin with the action of a particle in the geometrical form (the velocity of
light is equal to 1):

\begin{equation}
I=-m\int ds,  \label{1}
\end{equation}%
where $m$ is a mass of a particle, and

\begin{equation}
ds^{2}=dx^{2}\equiv dx^{\mu }dx_{\mu }  \label{2}
\end{equation}%
is the interval in the Minkowsky space. Here the Greece indices are lowered and
rised by means of metric tensor with the signature $\left( +,-,-,-\right) $.
Introducing an arbitrary parametrization of a world line, $x^{\mu }=x^{\mu
}\left( \tau \right) ,\tau \in \left[ 0,1\right] $, and defining a 4-vector
of the canonical momentum,

\begin{equation}
p_{\mu }\equiv -m\frac{\overset{\cdot }{x}_{\mu }}{\sqrt{\overset{\cdot }{x}%
^{2}}},  \label{3}
\end{equation}%
where the dot denotes the derivative with respect to the parameter $\tau $, one can write
action (\ref{1}) in the canonical form:

\begin{equation}
I=\int\limits_{0}^{1}\left( p_{\mu }{\overset{\cdot }{x}}^{\mu }-\chi
H\right) d\tau .  \label{4}
\end{equation}%
Here $\chi $ is a new variable which ensures the re-parametrization
invariance of the action (\ref{4}) and

\begin{equation}
H\equiv p^{2}-m^{2}.  \label{5}
\end{equation}%
At this stage an invariant parameter along a world line can be introduced:

\begin{equation}
c\left( \tau \right) =\int\limits_{0}^{\tau }\chi d\tau .  \label{6}
\end{equation}%
Then action (\ref{4}) takes a form:

\begin{equation}
I=\int\limits_{0}^{C}\left( p_{\mu }\overset{\cdot }{x}^{\mu }-H\right) dc,
\label{7}
\end{equation}%
where now the dot denotes the derivative with respect to the parameter $c$, and
$C\equiv c\left( 1\right)$.

Let us quantize action (\ref{7}) following \cite{GL}. In the space of
functionals $\Psi \left[ x^{\mu }\left( c\right) \right] $ we define a
functional-differential operator:

\begin{equation}
\widehat{p}_{\mu }\equiv \frac{\widetilde{\hbar }}{i}\frac{\delta }{\delta
x^{\mu }\left( \tau \right) },  \label{8}
\end{equation}%
where $\widetilde{\hbar }$ is a constant with the dimensionality $Dj\cdot
s^{2}$. For an action operator $\widehat{I}$ which is obtained by
substitution of (\ref{8}) into (\ref{7}), we consider the following secular equation:

\begin{equation}
\widehat{I}\Psi \equiv \int\limits_{0}^{C}\left[ \frac{\widetilde{\hbar }}{i%
}\overset{\cdot }{x}^{\mu }\frac{\delta \Psi }{\delta x^{\mu }}+\widetilde{%
\hbar }^{2}\frac{\delta ^{2}\Psi }{\left( \delta x^{\mu }\right) ^{2}}%
+m^{2}\Psi \right] dc=\lambda \Psi .  \label{9}
\end{equation}%
We formulate the quantum action principle as a condition of the extremum of
this eigenvalue $\lambda $ with respect to a set of parameters which will be
defined below.

Let us introduce an exponential representation of the wave functional:

\begin{equation}
\Psi \left[ x^{\mu }\left( c\right) \right] =\exp \left( \frac{\widetilde{%
\hbar }}{i}\sigma \left[ x^{\mu }\left( c\right) \right] +r\left[ x^{\mu
}\left( c\right) \right] \right) ,  \label{10}
\end{equation}%
where real functionals $\sigma \left[ x^{\mu }\left( c\right) \right]$ and $r\left[
x^{\mu }\left( c\right) \right] $ are supposed to be analytical. The latter
means that they are representable by series in the degrees $x^{\mu }\left( c\right) $.
With account of exponential representation (\ref{10}), from Eq.(\ref{9}) one
obtains an eigenvalue,

\begin{eqnarray}
\lambda &=&\int\limits_{0}^{C}\left[ \overset{\cdot }{x}^{\mu }\frac{\delta
\sigma }{\delta x^{\mu }}-\left( \frac{\delta \sigma }{\delta x^{\mu }}%
\right) ^{2}\right.  \notag \\
&&\left. +\widetilde{\hbar }^{2}\left( \left( \frac{\delta r}{\delta x^{\mu }%
}\right) ^{2}+\frac{\delta ^{2}r}{\left( \delta x^{\mu }\right) ^{2}}\right) %
\right] dc+m^{2}C,  \label{11}
\end{eqnarray}%
and, in addition, a condition of its reality,

\begin{equation}
\int\limits_{0}^{C}\left[ \overset{\cdot }{x}^{\mu }\frac{\delta r}{\delta
x^{\mu }}-2\frac{\delta \sigma }{\delta x^{\mu }}\frac{\delta r}{\delta
x_{\mu }}-\frac{\delta ^{2}\sigma }{\left( \delta x^{\mu }\right) ^{2}}%
\right] dc=0.  \label{12}
\end{equation}

Representation (\ref{11}) is not final because eigenvalues must be independent
on a world line $x^{\mu }\left( c\right) $ except for boundary
points $b^{\mu }\equiv x^{\mu }\left( C\right)$ and $a^{\mu }\equiv x^{\mu
}\left( 0\right) $ which are supposed to be fixed in the action
principle. This demand imposes 
a set of differential equations on
coefficients of series which are represented by the functionals $\sigma \left[
x^{\mu }\left( c\right) \right]$ and $r\left[ x^{\mu }\left( c\right) \right] $.
A solution of this set of equations depends on initial values of these coefficients at
the moment $c=0$. Therefore, we obtain an eigenvalue $\lambda $ as a
function of initial data. It is this function that must be stationary
in the quantum action principle. The variable $C$ also must be added to the set
of variational parameters. In the next section, a quasi-classical
approach for the quantum action principle will be considered, and the
classical limit of the quantum action of a free relativistic particle will
be obtained.

\section{CLASSICAL\textbf{\ LIMIT OF QUANTUM ACTION}}

In the classical limit, the eigenvalue $\lambda $ of the quantum action,
Eq. (\ref{11}), is completely defined by the functional $\sigma \left[
x^{\mu }\left( c\right) \right] $. In the case of a free relativistic
particle considered here, one can take into account only integral
functionals in the following form:

\begin{equation}
\sigma \left[ x^{\mu }\left( c\right) \right] =\int\limits_{0}^{C}\left[
\sigma _{1\mu }\left( c\right) x^{\mu }\left( c\right) +\frac{1}{2}\sigma
_{2}\left( c\right) \left( x\right) ^{2}+...\right] dc.  \label{13}
\end{equation}%
In the classical limit, one can consider functionals which are not higher than
quadratic in $x^{\mu }\left( c\right) $. Substituting (\ref{13})
into Eq.(\ref{11}), after integration by parts one obtains the final form
of the eigenvalue $\lambda $,

\begin{equation}
\lambda =\left( \sigma _{1\mu }x^{\mu }+\frac{1}{2}\sigma _{2}\left(
x\right) ^{2}\right) _{0}^{C}-\int\limits_{0}^{C}\sigma _{2}^{2}dc+m^{2}C,
\label{14}
\end{equation}%
and the condition of its independence on a world line $x^{\mu }\left(
c\right) $ in terms of two differential equations,

\begin{equation}
\overset{\cdot }{\sigma }_{1\mu }+2\sigma _{2}\sigma _{1\mu }=0,  \label{15}
\end{equation}

\begin{equation}
\overset{\cdot }{\sigma }_{2}+2\sigma _{2}^{2}=0.  \label{16}
\end{equation}

A general solution of equations (\ref{15}) and (\ref{16}) is

\begin{equation}
\sigma _{1\mu }\left( c\right) =\frac{\sigma _{1\mu }^{\left( 0\right) }}{%
1+2\sigma _{2}^{\left( 0\right) }c},  \label{17}
\end{equation}

\begin{equation}
\sigma _{2}\left( c\right) =\frac{\sigma _{2}^{\left( 0\right) }}{1+2\sigma
_{2}^{\left( 0\right) }c},  \label{18}
\end{equation}%
where $\sigma _{1\mu }^{\left( 0\right) }$ and $\sigma _{2}^{\left( 0\right) }$
are initial values of the coefficients $\sigma _{1\mu }$ and $\sigma _{2}$.
Substitution of this solution into Eq.(\ref{14}) gives the eigenvalue $%
\lambda $ as a function of the initial data, $\sigma _{1\mu }^{\left(
0\right) }$, $\sigma _{2}^{\left( 0\right) }$, and the invariant time parameter $%
C$:

\begin{eqnarray}
\lambda &=&\sigma _{1\mu }^{\left( 0\right) }\left( \frac{b^{\mu }}{%
1+2\sigma _{2}^{\left( 0\right) }C}-a^{\mu }\right) +\frac{\sigma
_{2}^{\left( 0\right) }}{2}\left( \frac{\left( b^{\mu }\right) ^{2}}{%
1+2\sigma _{2}^{\left( 0\right) }C}\right.  \notag \\
&&\left. -\left( a^{\mu }\right) ^{2}\right) -\frac{\left( \sigma _{1\mu
}^{\left( 0\right) }\right) ^{2}}{1+2\sigma _{2}^{\left( 0\right) }C}+m^{2}C.
\label{19}
\end{eqnarray}%
It is this function that must be stationary with respect to the initial data, $%
\sigma _{1\mu }^{\left( 0\right) }$, $\sigma _{2}^{\left( 0\right) }$, and the
invariant time parameter $C$ in the quantum action principle. The extremum
condition with respect to the initial data gives

\begin{equation}
\sigma _{1\mu }^{\left( 0\right) }=\frac{1}{2C}\left[ b_{\mu }-a_{\mu
}\left( 1+2\sigma _{2}^{\left( 0\right) }C\right) \right] .  \label{20}
\end{equation}%
Therefore, one of the initial data parameters, $\sigma _{2}^{\left( 0\right)
}$ in this case, is not fixed in the classical limit of the quantum action
principle, and the eigenvalue $\lambda $ is degenerate. The extremum
condition with respect to $C$ gives

\begin{equation}
C=\pm \frac{1}{2m}\sqrt{\left( b-a\right) ^{2}}.  \label{21}
\end{equation}%
Substituting (\ref{20}), (\ref{21}) into Eq.(\ref{19}), one obtains the
quantum action eigenvalue in the classical limit:

\begin{equation}
\lambda =\pm m\sqrt{\left( b-a\right) ^{2}}.  \label{22}
\end{equation}%
This result coincides with classical action (\ref{1}) calculated on the
classical trajectory of a free particle. The wave functional corresponding
to eigenvalue (\ref{22}) has a phase which in the classical limit is
proportional to:

\begin{equation}
\sigma \left[ x^{\mu }\right] =\frac{1}{4}\int\limits_{0}^{Q}\left( x^{\mu
}\left( q\right) -\widetilde{x}\right) ^{2}dq,  \label{23}
\end{equation}%
where

\begin{equation}
\widetilde{x}^{\mu }\equiv -\frac{b^{\mu }-e^{Q}a^{\mu }}{e^{Q}-1},
\label{24}
\end{equation}
\begin{equation}
Q\equiv \ln \left( 1+2\sigma _{2}^{\left( 0\right) }C\right) .  \label{25}
\end{equation}

\section{\textbf{CONCLUSIONS}}

We conclude that in the classical limit the quantum action principle returns us to
the original action of a relativistic particle calculated on a classical
trajectory. Quantum corrections to this action will give us essential
predictions of the new theory and define a new "Plank" constant $\widetilde{%
\hbar }$. The parameter $\sigma _{1}^{\left( 0\right) }$, which is
indefinite in the classical limit, plays the role of a degree of excitation
of a quantum particle.

\begin{acknowledgments}

We are thanks V. A. Franke and A. V. Goltsev for useful discussions.

\end{acknowledgments}




\end{document}